# Single molecule narrowfield microscopy of protein-DNA binding dynamics in glucose signal transduction of live yeast cells

*Adam J.M. Wollman[1,2] and Mark C. Leake[1]*

[1] Biological Physical Sciences Institute (BPSI)

University of York

York

YO10 5DD

United Kingdom

[2] Corresponding author

e–mail: adam.wollman@york.ac.uk

Tel: +44 (0)1904 322697

Running head: Single-molecule narrowfield.

# Abstract

Single-molecule narrowfield microscopy is a versatile tool to investigate a diverse range of protein dynamics in live cells and has been extensively used in bacteria. Here, we describe how these methods can be extended to larger eukaryotic, yeast cells, which contain sub-cellular compartments. We describe how to obtain single-molecule microscopy data but also how to analyse these data to track and obtain the stoichiometry of molecular complexes diffusing in the cell. We chose glucose mediated signal transduction of live yeast cells as the system to demonstrate these single-molecule techniques as transcriptional regulation is fundamentally a single molecule problem – a single repressor protein binding a single binding site in the genome can dramatically alter behaviour at the whole cell and population level.

## 1. Introduction

Bulk biochemical methods can only measure mean ensemble properties while single molecule techniques allow the heterogeneity in molecular biology to be explored which often leads to a new understanding of the biological system involved.*(1)* The use of fluorescent protein fusions to act as reporters can provide significant insight into a wide range of biological processes and molecular machines, for enabling insight into stoichiometry and architecture as well as details of molecular mobility inside living, functional cells with their native physiological context intact.*(2–7)* Single-molecule narrowfield microscopy, and its similar counterpart Slimfield microscopy, is a versatile tool to investigate a diverse range of protein dynamics in live cells which can be used in conjunction with fluorescent protein fusion strains to generate enormous insight into biological processes at the single-molecule level. In bacteria, it has been used to investigate the components of the replisome*(8)* and the structural maintenance of chromosomes.*(9)*

In narrowfield microscopy, the normal fluorescence excitation field is reduced to encompass only a single cell. This produces a Gaussian excitation field (~30 µm$^2$) with 100–1000 times the laser excitation intensity of standard epifluorescence microscopy. This intense illumination causes fluorophores to emit many more photons, generating much greater signal intensity relative to normal camera-imaging noise and, hence, facilitates millisecond time-scale imaging of single fluorescently-labelled proteins. The millisecond time scale is fast enough to keep up with the diffusional motion present in the cytoplasm of cells and can also sample the fast molecular transitions that occur, particularly during signal transduction. Single fluorescent proteins or complexes of proteins can be considered point sources of light and so appear as spatially extended spots in a fluorescence image due to diffraction by the microscope optics.*(10)* Narrowfield microscopy data consists of a time-series of images of spots which require a significant amount of *in silico* analysis. Spots must be identified by software, the intensity of these spots quantified to calculate their stoichiometry and their position tracked over time to produce a trajectory.

We have applied narrowfield microscopy to glucose signal transduction in budding yeast, *Saccharomyces cerevisiae*. All cells dynamically sense their environment through signal transduction mechanisms. The majority of these mechanisms rely on gene regulation through cascades of protein-protein interactions which transmit signals from sensory elements to responsive elements within each cell. The Mig1 protein is an essential transcription factor in this mechanism in yeast. Mig1 is a Cys2-His2 zinc finger DNA binding protein*(11)* which binds several glucose-repressed promoters.*(12–15)* In the presence of extracellular glucose it is poorly phosphorylated and predominantly located in the nucleus*(16, 17)* where it recruits a repression complex to the DNA.*(18)* If extracellular glucose concentrations levels are depleted, Mig1 is phosphorylated by the sucrose non-fermenting protein (Snf1),*(19–21)* resulting in a redistribution of mean localization of Mig1 into the cytoplasm.*(16, 22, 23)* Thus, Mig1 concentration levels in the cell nucleus and cytoplasm serve as a readout of glucose signal transduction in budding yeast.*(24)* Mig1 has been labelled with the green fluorescent protein, GFP, and in the same strain, a ribosome component, Nrd1, almost completely localised to the nucleus, has been labelled with the mCherry fluorescent protein.*(17)* We have used narrowfield microscopy to track single Mig1-GFP complexes as they diffuse in the nucleus and cytoplasm in the presence and absence of extra-cellular glucose.

Here we describe in detail how to obtain single molecule data of fluorescently labelled transcription factors in live yeast but also methods used to analyse the data obtained. We describe ADEMS code,*(25)* the custom Matlab software we have created to track fluorescent molecules and quantify their stoichiometry. We also show how fluorescence images of Mig1-GFP and Nrd1-mcherry can be segmented to identify the boundary of the cell and nucleus respectively and thus how trajectories' can be categorised by their different sub-cellular compartments.

## 2. Materials

### 2.1 Fluorescently labelled yeast strains

1. MATa MIG1-GFPHIS3 NRD1-mCherry- hphNT1METLYS S. cerivisiae strain in the BY4741 background*(17)* stored at -80˚C in YNB media supplemented with 20% glycerol.
2. Yeast Nitrogen Base (YNB) media (Formedium) pH 6

## 2.2 Sample Preparation

1. Standard Microscope slides (Fisher)
2. Coverslips (Menzel-Gläser)
3. Gene frames (17mm x 28mm) and spreaders (Thermo Scientific)
4. 2% Agarose solution
5. Plasma cleaner (Harrick Plasmas PDC-32G)
6. Desk top centrifuge (Sigma 1-14)

## 2.3 Narrowfield Microscope

For narrowfield microscopy, ~6 W/cm$^2$ excitation light must be delivered to the sample centred at 488 nm for millisecond imaging of GFP. This must be combined with a high speed (kHz) camera capable of detecting single fluorescent proteins. Our microscope is constructed from:

1. A Zeiss microscope body with a 100x TIRF 1.49 numerical aperture (NA) Olympus oil immersion objective lens and an xyz nano positioning stage (Nanodrive, Mad City Labs).
2. 50 mW Obis 488 nm and 561 nm lasers for fluorescence excitation.
3. A dual pass GFP/mCherry dichroic with 25nm transmission windows centred on 525nm and 625nm was used underneath the objective lens turret.
4. A high speed camera (iXon DV860-BI, Andor Technology, UK) was used to image at typically 5ms/frame with the magnification set at ~80 nm per pixel.
5. The camera CCD was split between a GFP and mCherry channel using a bespoke colour splitter consisting of a dichroic centred at pass wavelength 560 nm and emission filters with 25 nm bandwidths centred at 525 nm and 594 nm.

6. The microscope was controlled using our in-house bespoke LabVIEW (National Instruments) software.

## 2.4 Computational Analysis

1. We use MATLAB 2014b, which has the advantage that many functions are built in such as curve fitting etc. but these methods could be implemented in other packages such as IDL, LabVIEW or Python, C++ or Java.

# 3. Methods

## 3.1 Growing yeast strains

1. Prepare 5mL YNB supplemented with 4% Glucose in 15mL Falcon tube
2. Scrape a small quantity of frozen yeast culture from the cryovial using a pipette tip without defrosting the vial.
3. Swirl pipette tip in prepared media and leave the culture to grow overnight at 30°C, shaking at 200rpm.

## 3.2 Plasma cleaning coverslips

Glass coverslips must be plasma cleaned before use to remove any material left on the glass from manufacture. We have found this material to be fluorescent under the intense excitation light of narrowfield microscopy. This produces a fluorescent background in an image or false positive spots.

1. Place coverslips into the plasma cleaner
2. Seal with the valve door and turn on the vacuum pump, pressing the edges of the door to insure a good seal.
3. Turn the radio frequency (RF) generator to high. If the vacuum pressure is low enough, a pink-violet glow will emanate from the chamber, but if the pressure is not yet low enough, continue to press on the seal until more air is evacuated by the pump.
4. Once plasma is generated, slightly open the valve to allow a small amount of air into the chamber. The plasma will change colour to violet indicating oxygen plasma.
5. After coverslips have been exposed to oxygen plasma for ~1 min, turn off the RF and vacuum pump and slowly open the valve.
6. Once atmospheric pressure is restored, remove the coverslips and store in a clean petri dish.

### 3.3 Preparing cell samples

1. For glucose conditions, the overnight culture can be used as the cell sample.
2. For absence of glucose conditions, prepare 1 mL of YNB.
3. Pellet 1 mL of cell culture by spinning in a desktop centrifuge for 2mins at 3000rpm and discard the supernatant
4. Re-suspend in 500 µl YNB and pellet again.
5. Discard the supernatant and re-suspend in 100 µl YNB or more depending on cell density.

### 3.4 Agarose pad preparation

Cells were imaged on agarose pads suffused with media. This ensures cells remain healthy during imaging but also immobilises them. Figure 1 illustrates agarose pad assembly

1. Prepare 500 μl 2xYNB, supplemented with 8% glucose to image cells in high glucose conditions.
2. Remove the larger of the two clear plastic covers from the gene frame and apply the frame to a glass slide to create a rectangular well.
3. Melt 2% agarose solution in a microwave
4. Pipette 500 μl of hot agarose into the 2xYNB and quickly mix before removing 500 μl and pipetting into the well on the slide.
5. Quickly apply a plastic spreader (included with Gene Frames) to the well to remove excess agarose and leave a thin even layer in the well.
6. Slide the spreader off the pad carefully and remove the second plastic cover from the gene frame.
7. Pipette 5ul of cell sample onto the pad in ~10 droplets and leave to dry for ~5 mins before covering with a plasma cleaned coverslip.

### 3.5 Obtaining single molecule data

1. Place the sample on the microscope and locate a cell by imaging in brightfield.
2. Focus on the mid body of the cell by adjusting the focus to the point of minimum contrast.
3. Acquire 10 brightfield frames at 50ms exposure time with no camera gain. All images saved with raw pixel values as stacked Tag Image Format (TIF) files.
4. Turn off the brightfield and set gain to maximum.
5. Acquire 100 frames at 5 ms exposure time with the 561 nm laser at 15 mW power to obtain images of the mCherry signal, this will be used to define the nucleus.
6. Acquire 1000 frames at 5 ms exposure time with the 488 nm laser at 30 mW power to obtain a time series of GFP fluorescence images. (see Note 1)

7. Repeat for ~30 cells or however many are required for a robust statistical sample. (see Note 2)

### 3.6 Tracking single molecules

Bespoke Matlab code has been written to track bright spots in fluorescence image time series. The steps performed by the code are outlined here.

1. Load TIF file containing single molecule tracks, in this case the 1000 frame 488 nm exposure, into MATLAB as an *m x n x p* array, *m* pixels by *n* pixels by *p* frames.
2. Apply a top hat transformation to each frame to even the background and threshold the resulting image using Otsu's method.
3. Dilate the resulting binary image with a disk shaped structural element and then erode with the same element to remove bright noise pixels.
4. Perform an ultimate erosion to leave only non-zero pixels at potential spot locations. (see Note 3)
5. At each of these potential spot locations, define a square 16 pixel region of interest (ROI) and a 5 pixel radius circular ROI centred on the intensity centroid of the potential spot.
6. Convolve the circular ROI with a 2D Gaussian function with 3 pixel width centred on the current intensity centroid and use this to determine a new intensity centroid
7. Repeat step 6 until the centroid position converges on the final sub-pixel spot centre coordinates. Figure 2 illustrates one iteration of a Mig1-GFP spot.
8. Define the spot's total intensity as the sum of the pixel values inside the circular ROI corrected for background by subtracting the mean pixel value of the remaining pixels in the square ROI.

9. Define the spots signal to noise ratio (SNR) as the spot's total intensity divided by the standard deviation of the remaining pixels in the square ROI and discard spots with SNR<0.4.
10. The centre of the mask gives the sub-pixel centroid coordinates of the spot.
11. Determine the width of each spot by constrained fitting of a 2D Gaussian function inside the square ROI, with the width and central intensity the only variables.
12. Once all spots are found and characterised in at least 2 consecutive frames, they can be linked together into trajectories.
13. Calculate the pairwise distance between all pairs of spots in consecutive frames and keep any which are below 5 pixels.
14. Link closest pairs as long as their intensities or widths do not differ by >2x and assign a new trajectory number or continue an existing one if a spot is already part of a trajectory. Thus a complete set of trajectories for the time series is acquired.

## 3.7 Analysing single molecule trajectories

Once trajectories are obtained, they are analysed to determine the number of fluorophores present in each spot. The intensity of a single fluorophore is first found from the distribution of all spot intensities over the whole time series. The most common intensity value is that of a single fluorophore as all traces bleach to this value. The number of fluorophores present in each spot can be determined by dividing the initial value by the single fluorophore value. The initial intensity is determined by fitting an exponential decay function to each trace.

1. Generate histogram or kernel density estimation (KDE) of all the intensity values of all the spots found in trajectories. Figure 3 left shows the KDE of mig1-GFP intensity values obtained in a single cell.
2. As every complex of fluorophores is bleaching, the most common intensity value in the intensity distribution is the characteristic single fluorophore intensity and the peak value in the distribution. Figure 3 right shows spot intensity as a function of time with the single GFP value marked with a line. (see Note 4)

3. Fit an exponential to all the spot intensity values as a function of time.
4. Use the time constant from the global fit, to fit exponentials to each trajectory, provided it is >3 points, within the first 200 frames and its initial intensity is >2x the single fluorophore intensity.
5. The number of fluorophores present in each spot is the initial intensity in each trajectory's fit divided by the characteristic single fluorophore intensity. (see Note 5)

### 3.8 Categorizing tracks by cell compartment

Trajectories are analysed in terms of their location in the cell. The cell and nuclear boundaries are determined by segmenting GFP and mCherry fluorescence frame averaged images. This allows trajectories to be defined as nuclear, cytoplasmic and trans-nuclear. Figure 4 shows example frame averages and segmentation and categorised trajectories overlaid on a brightfield image of a yeast cell.

1. Generate frame averages over first five frames of the Nrd1-mCherry and Mig1-GFP acquisitions. These will be used to determine the nucleus and cell boundary.
2. Threshold these images above the full width half maximum (FWHM) of the background peak in the pixel intensity distribution to obtain a binary image.
3. Erode this image with a 4 pixel disk shaped structural element to remove any bright single pixels and create a smoother edge around the object. These masks define the nucleus and cell pixels.
4. Divide the tracks into those which are always in the nucleus, always in the cytoplasm or are in both at different times - trans-nuclear tracks.

## 4. Notes

1. During fluorescence acquisitions, the camera must begin acquiring frames just before the laser illuminates the sample. This allows all of the emitted light from the sample to be captured by the camera which is crucial for analysis.
2. When moving through the sample during imaging, care must be taken to move in a set direction through the sample. This ensures that every cell that is imaged has not had any prior exposure to the laser and thus no photobleaching has occurred.
3. Steps 2-4 in section 3.6 identify candidate spots in fluorescence images. This process is not strictly necessary as spots are evaluated using iterative Gaussian masking which could be performed at every pixel location in an image. This would be very computationally intensive but by identifying candidates, the number of possible spot locations is reduced along with the processing time.
4. The intensity of a single fluorophore can be independently verified in an *in vitro* assay by imaging antibody immobilised fluorophores on a coverslip. This intensity will vary from that observed *in vivo* due to the different conditions inside a cell, particularly the differing pH.
5. Any fluorophores which are within the diffraction limited spot width of each other (~250 nm for GFP) will appear to be a single spot in a narrowfield image. This distance is large compared to the size of proteins such as Mig1 of a few nanometres and so it is possible that some spots in a frame are not molecular complexes but separate molecules within 250nm of each other. These events are short lived for diffusing molecules and so the stoichiometry of complexes is obtained from the fits to the intensity vs time traces.

## Acknowledgments

We thank Sviatlana Shashkova and Stefan Hohmann (University of Gothenburg, Sweden) for donation of yeast cell strains and assistance with yeast cell culturing. MCL was assisted by a Royal Society URF and research funds from the Biological Physical Sciences Institute (BPSI) of the University of York, UK.

**Figure legends:**

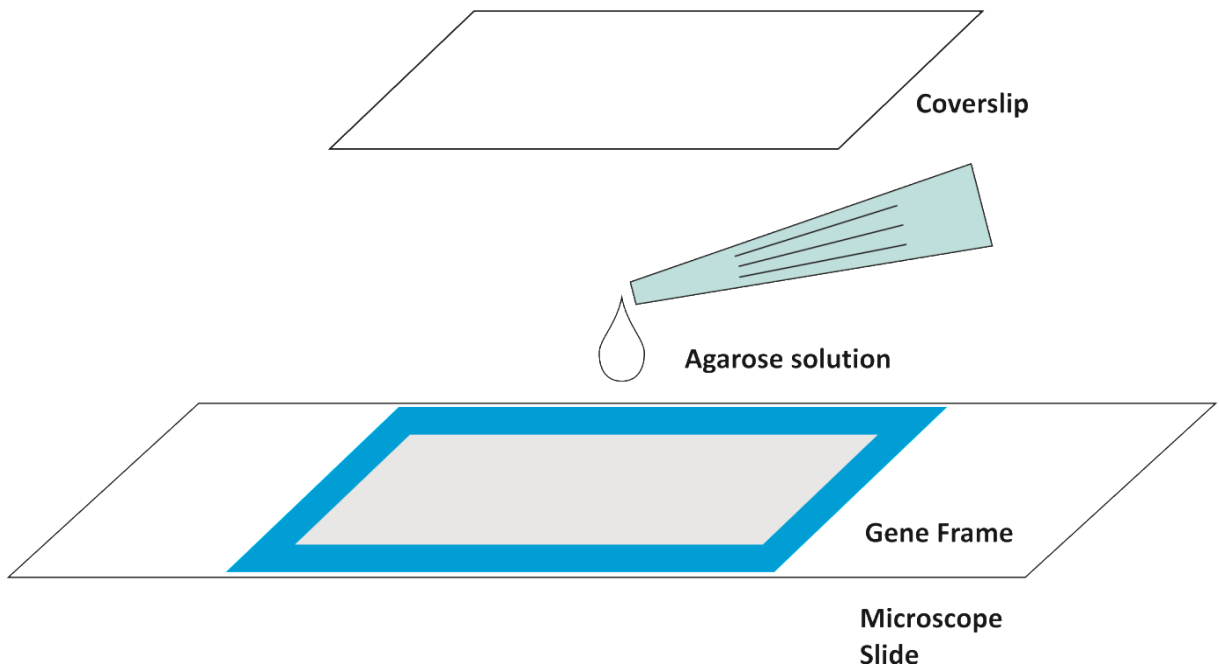

Figure 1: Schematic of agarose pad assembly

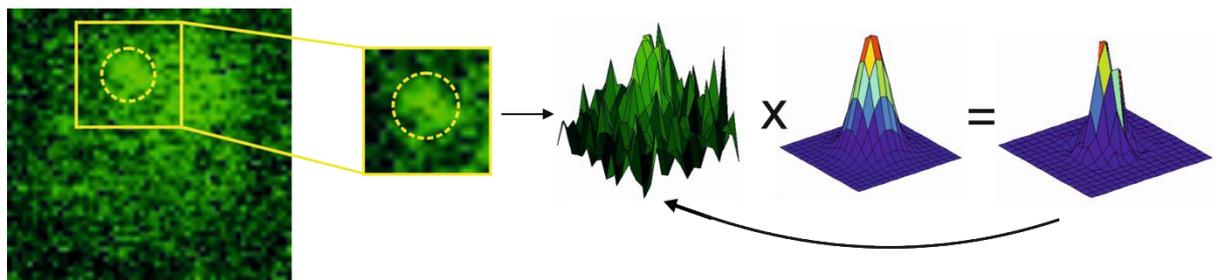

Figure 2: Schematic of iterative Gaussian masking to determine spot centroid. Left: a fluorescence image of Mig1-GFP in a yeast cell with a spot identifiable. Right: schematic illustration of iterative Gaussian masking to find the spot centroid. The spot pixels are shown as a 3D surface and convolved with a 2D Gaussian centred on the centroid estimate. The resulting image is used to find a new centroid estimate and the process iterated until convergence.

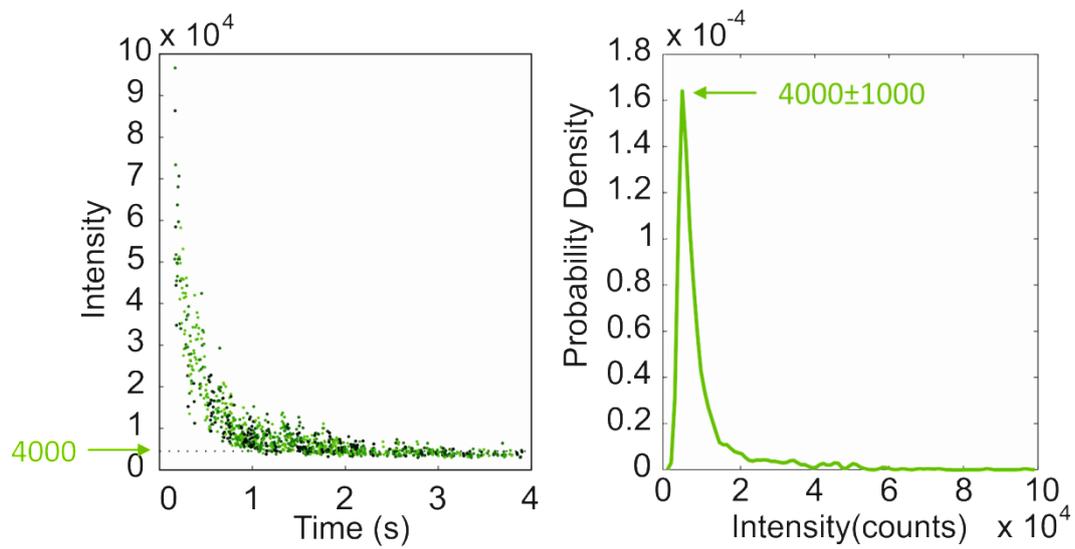

Figure 3: Left spot intensity as a function of time with the single GFP intensity value marked as a dotted line. Right distribution of spot intensities with the single GFP value marked.

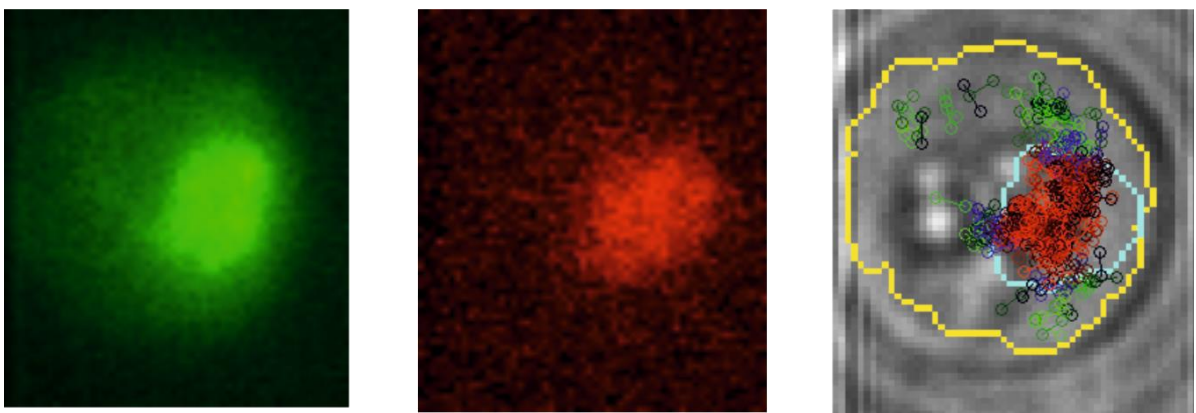

Figure 4: Left fluorescence image of Mig1-GFP, Middle fluorescence image of Nrd1-mCherry, Right brightfield image with cell and nucleus outlines (yellow and cyan respectively) and nuclear, cytoplasmic and trans-nuclear tracks (red, green and blue repsectively) overlayed.